\documentclass[12pt]{article}
\usepackage{amsmath,amsfonts,amssymb,amsthm}
\usepackage{mathtext}
\usepackage{epsfig,enumerate}

\newtheorem*{thm}{Theorem}

\begin{document}

\centerline{\bf P.Grinevich, S.Novikov\footnote {P.G.Grinevich,
 Landau Institute for Theoretical Physics, Moscow, Russia\\
e-mail pgg@landau.ac.ru,\\ S.P.Novikov, University of Maryland
at College Park, USA, and Landau/Steklov Institutes
 for Theoretical Physics and Mathematics, Moscow,
 Russia\\ e-mail novikov@umd.edu}$^,$}

\vspace{0.5cm}

\centerline{\Large On the s-meromorphic OD operators}

\vspace{1cm}

{\LARGE Abstract}. {\it We consider
linear spectral-meromorphic (s-meromorphic) OD operators at the real axis such that all local solutions to the eigenvalue
problems are meromorphic for all $\lambda$. By definition,  rank one algebro-geometrical
operator $L$  admit an OD operator $A$ such that $[L,A]=0$ and rank of this commuting pair is equal to one. All of them
are s-meromorphic.
In particular, second order ``singular soliton'' operators satisfy to this condition. Operator $L^+$
formally  adjoint to
s-meromorphic operator $L$ is also s-meromorphic. For singular eigenfunctions of operators $L,L^+$ following scalar product  $<f,g>=\int_{R} f\bar{g}dx$ is well-defined such that $<Lf,g>=<f,L^+g>$ avoiding isolated singular points.  For the case $L=L^+$ this formula defines indefinite inner product on the spaces of singular functions $f,g\in F_L$ associated with operator $L$. They are $C^{\infty}$ outside of singularities  and have isolated singularities of the same type as eigenfunctions $Lf=\lambda f$.  Every s-meromorphic operator can be approximated by
algebro-geometric rank one operators in any finite interval.}

\pagebreak

{\bf Definitions}.{\it  1.We call linear OD operator $L$ algebrogeometric if there exist a nontrivial OD operator $A$ with property $[L,A]=0$.
According to the classical (Burchnall-Chaundy) lemma, there exists polynomial such that $P(L,A)=0$ defining the Burchnall-Chaundy (BC) algebraic curve $\Gamma$ where $P(\lambda,\mu)=0$. For every point $\gamma\in\Gamma$ we have $k$-dimensional linear space  of common eigenfunctions of operators $L$ and $A$. Here $k$ is divisor of both orders of operators, not necessarily maximal. For the relatively prime orders we always  have $k=1$.
2.We call $k$ the {\bf rank} of operator $L$ and of the commuting pair $A,L$.}

 Commuting operators of relatively prime orders were classified in 1920-s by Burchnall and Chaundy \cite{BC1}, \cite{BC2}. Their  ideas were discussed also by A.Baker in 1920s  but no development followed. This theory was finally completed much later by Krichever \cite{K1} after the great development of Periodic
Soliton Theory (or Periodic Analog of Inverse Scattering Transform) in Mathematical Physics {\cite{N74,DMN76}. This theory was created to find analytically a broad family of exact ''algebrogeometric'' or finite-gap solutions with periodic boundary conditions to the famous KdV equation in 1974. It was extended to other famous equations like Nonlinear Schrodinger, Sine-Gordon and especially for 2+1 systems like KP (\cite{K1}. Soliton Theory produced also a far reaching development of the spectral theory of periodic Schrodinger operator. The first examples of (nonsingular) finite-gap periodic potentials were discovered  in 1950s studying the ''shifted Lame' potentials'' by Ince and others--see the book of Magnus and Winkler \cite{MW}). They  were classified and effectively calculated in the Theory of Solitons in 1974. This theory appeared completely independently from the algebraic ideas \cite{BC1},  \cite{BC2} discovered in 1920s. Only in 1970s theta-functional formulas were finally obtained for finite-gap potentials and KdV solutions, and later also for eigenfunctions of the generic rank one algebrogeometric operators.

 Soliton Theory allowed also
to make breakthrough in the classical Burchnall-Chaundy Problem- how to  classify and calculate the higher rank commuting operators $k>1$.
It was pointed out already in \cite{BC1}, \cite{BC2} that this problem is very difficult. Misterious examples of the higher rank operators with polynomial coefficients were found by Dixmier in 1960s for the nonsingular genus one
BC-curves $\Gamma$. Nothing like that is possible for the rank one case where nonsingular curves always lead to abelian varieties and therefore to quasiperiodic coefficients. Modern mathematics associates the higher rank  problem with theory of holomorphic vector bundles over Riemann Surfaces.  In the Soliton Theory this problem is related also to the famous 2+1 KP system. New methods were developed and  effective analytical  calculation of coefficients of operators for nonsingular genus one curve were obtained by Novikov and Krichever \cite{KN1}. For rank 2 and more eigenfunctions of algebrogeometric  operators in general cannot be explicitly calculated.
Important calculations of operators with polynomial coefficients were performed for genus one by Grinevich \cite{G1} and Mokhov \cite{Mo}. Modern development for genus higher than one started after the works of Mironov \cite{Mir}.

{\bf In the present work we are going to work only with rank one algebrogeometric operators.}

{\bf Definitions}. 3.We call operator with isolated singularities at the real $x$ {\bf spectral-meromorphic or s-meromorphic} if all eigenfunctions $Lf=\lambda f$ are meromorphic nearby of singular points.

It follows from the periodic Soliton Theory that all rank one algebrogeometric operators are s-meromorphic. Eigenfunctions correspond to the data on the Riemann surfaces. They are always meromorphic even if data lead to the singular operators with poles.

 For the formally self-adjoint case we proved recently for all higher order algebrogeometric rank one operators
\cite{GN4} that product of 2 eigenfunctions $fg$ never has nontrivial Laurent  terms of the order $(x-x_j)^{-1}$ at any singular point $x_j$.

 For the second order operators $L=-\partial^2+u(x)$ this observation played key role  in our series of works from the very beginning (see \cite{GN1}. The singularity of s-meromorphic second order operator looks like $$u=n(n+1)y^{-2} +\sum_{ 0\leq  q< n}c_qy^{2q}+O(y^{2n}), y=(x-x_j) $$
 $$f=a_0y^{-n}+a_1y^{n-2}+...+a_ky^{n-2k}+...+a_{n}y^{n}+O(y^{n+1}),\\\\\ Lf=\lambda f$$

In the work \cite{APP} this condition was found for the simplest case $u=2/y^2 +...$ as necessary for the definition of the scattering coefficients in the rapidly decreasing singular case. For rational KdV solutions this condition was  found in \cite{DuGr}. It is true for all algebrogeometric solutions as we know. All rational solutions indeed are algebrogeometric.
{\bf Question: Consider the general KdV solutions $u(x,t)$  $x$-meromorphic for every moment of time. Are potentials $u(x,t)$ always s-meromorphic in this case?}

 For the higher orders we proved   property of the product $fg$ only recently \cite{GN3}). It leads to conclusion that
following formula defines correctly the inner product $$<f,g>=\int_{<R>}f(x)\bar{g}(\bar{x}))dx$$
Here sign $<R>$ means that we are
integrating along the real line between critical points avoiding them from upper or lower site. Our condition leads to conclusion that result does not depend on path. This inner product is indefinite. Number of negative squares is an integral of time dynamics for KdV hierarchy if order is 2,
and Gelfand-Dikii hierarchies\cite{GD} for the case of higher order operators $L$.

For nonselfadjoint algebrogeometric rank one operators everything is the same but scalar product under discussion is well-defined between
the eigenfunctions of operators $L$ and $L^+$. Both operators corresponds to the same Riemann surface but have different divisors of poles $D,D^+$
where $D+D^+\sim K+ 2P$, $P$ is the infinite point.
It means generically that they  have the same type generic  singularities at the axis $R$ . We call these generic singularities the singularities associated with given Riemann surface.

Consider any discrete set $X$ of isolated point $x_j\in R$ and fix some finite-dimensional linear subspaces $
L_j$ in the space of polynomials from the negative powers of  the variable $y=x-x_j$ without zero term for every point $x_j$. Define space $F_{X,L}$ of such functions that

1.They are $C^{\infty}$ outside of the points $x_j$

2.They are meromorphic in the small vicinity of every point $x_j$

3.Their principal parts at the points $x_j$ belong to the subspaces $L_j$

4.Product of functions do not have terms of the order $y^{-1}$ in the singular points $x_j,
 y=x-x_j$. This requirement leads to the restrictions on the positive parts of functions at the points $x_j$.

Following our previous works \cite{GN1,GN2,GN3,GN4}, we define inner product in the space $F_{X,L}$ as above
$$<f,g>=\int_{<R>}f\bar{g}(\bar{x})dx$$ avoiding singularities from upper or lower sides.

Such spaces $L_j$ can be easily classified. In every such space a {\bf Canonical triangular basis} $f_1,...,f_l$ can be chosen
corresponding to the number of negative integers $-n_1<n_2,...<-n_l<0$. The basic elements has a form
$$f_k=y^{-n_k}+\sum_{j}\alpha_{k,j}y^{-m_{k,j}}$$ where $m_{k,j}<n_k$ and  $m_{k,j}\neq n_s$ for any $s,k$.

For the second order operators $L=-\partial^2+u(x)$ the spaces $L_j$ consist from the following powers starting from some integer even or odd: $$n_1=2l+1,n_j=n_1-2j,...,n_{l+1}=1$$
or $$n_1=2l,...,n_l=2$$

In both cases dimension of this spaces  is equal to the number of negative squares for the inner product
contributed by one singular point.

The important property of the s-meromorphic operators is the correct analytical definition of the transition matrix between any pair of nonsingular points $T_{x_0,x_1}=(t_{ij})$
defined through the basic solutions $Lf_i^{x_0}=\lambda f_i^{x_0}$ such that $f_i^{k}|_{x=x_0}=\delta_i^k, i,k=0,1,...,p-1$ where $p$ is the order of $L$
We define as usual the transition matrix as $$f_i^{x_0}=\sum_jt_{ij}^{(x_0,x_1)}f_j^{x_1}$$

All solutions are one-valued and $x$-meromorphic in the singular points. Everything is analytical in $\lambda$.
For $T$-periodic operators we define {\bf Monodromy Matrix } where $x_1=x_0+T$

Therefore  {\bf Riemann Surface} of the Bloch-Floquet eigenfunction is well-defined as in the nonsingular case. Its properties
for $\lambda\rightarrow \infty$ are completely similar to the nonsingular case but this statement needs full rigorous proof.
Closing branching points nearby of infinity, we are coming to finite genus and therefore to the following result:

\begin{thm}For any finite interval every s-meromorphic periodic (as well as nonperiodic) operator $L$ can be approximated by the algebrogeometric rank one operator
\end{thm}

In process of approximation all discrete invariants of singular points (see above) remain unchanged.

{\bf Problem: Which subspaces   $L_j$ above can be realized by the s-meromorphic operator.}

\end{document}